\documentclass[11pt]{article} 
\usepackage{cite}
\usepackage{latexsym}
\usepackage{amssymb}
\usepackage{amsmath}
\usepackage[breaklinks=true,linktocpage=true]{hyperref}
\hypersetup{colorlinks=true, allcolors=blue}
\usepackage{graphicx}

\usepackage{slashed}
\usepackage{mathbbol}

\usepackage{color}
\usepackage{comment}
\usepackage{physics}
\usepackage{float}
\usepackage{siunitx}

\usepackage{multicol}

\begin{document}
\pagestyle{empty}
\begin{flushright}
Nufact 2024-06
\end{flushright}
\begin{center}
{\bf\LARGE
T violation at a future neutrino factory
}
\\
\vspace*{0.5cm}
 Contribution to the 25th International Workshop on Neutrinos from Accelerators
 \\

\vspace*{0.5cm}
{\large 
Ryuichiro Kitano$^{1,2}$, 
Joe Sato$^{3}$, and 
Sho Sugama$^{3}$
} \\
\vspace*{0.5cm}

{\it 
$^1$KEK Theory Center, Tsukuba 305-0801,
Japan\\
$^2$Graduate University for Advanced Studies (Sokendai), Tsukuba
305-0801, Japan\\
$^3$Department of Physics, Facility of Engineering Science, Yokohama National University, Yokohama 240-8501, Japan 
}

\end{center}


\begin{abstract}
{\normalsize
We study the possibility of measuring T (time reversal) violation
in a future long
baseline neutrino oscillation experiment. By assuming a neutrino
factory as a staging scenario of a muon collider at the J-PARC site,
we find that the $\nu_e \to \nu_\mu$ oscillation probabilities can be
measured with good accuracy at the Hyper-Kamiokande detector. 
By comparing with the probability of the time-reversal process,
$\nu_\mu \to \nu_e$, measured at the T2K/T2HK \cite{T2K:2021xwb,Hyper-Kamiokande:2018ofw}, one can
determine the CP phase $\delta$ in the neutrino mixing matrix if
$|\sin(\delta)|$ is large enough. 
The determination of $\delta$ can be made with poor knowledge of the
matter density of the earth as T violation is almost insensitive to the
matter effects. The comparison of CP and T-violation measurements,
{\it \`a la} the CPT theorem, provides us with a non-trivial check of
the three neutrino paradigm based on the quantum field theory.
This proceeding is based on JHEP \textbf{12} (2024), 014 [arXiv:hep-ph/2407.05807].
}
\end{abstract} 

\baselineskip=18pt
\begin{multicols}{2}
%
In the current situation, the discovery of CP violation in
the neutrino sector itself will be the most important milestone. However, imperfect knowledge of the matter-density profile of the earth can fake the CP violation.
In contrast, T violation is insensitive to the matter
effects.
The importance of measuring T violation has been pointed out by many physicists. And many of them considered 
the use of high intensity muon beams, so called 
neutrino factory~\cite{Geer:1997iz}. It is also interesting to compare CP-violation measurements with
T-violation ones. Recently, a muon collider was proposed in Japan~\cite{Hamada:2022mua}.
For example, as one of the staging scenarios, one can consider a low energy $\mu^+$
beam such as $O(1)$~GeV from the neutrino factory to shoot $\nu_e$ beams to
Super/Hyper-Kamiokande.
In this paper, we study the possibility of measuring T violation. As an example, we take the same
baseline as the T2HK experiments~\cite{Hyper-Kamiokande:2018ofw}.
\label{sec:chi2}

We discuss the statistical precisions of the CP and T violation
measurement. We assume a race-track type 
storage ring of the  high-intensity muon beam at the J-PARC site with one of the straight section pointing towards the Hyper-Kamiokande detector.
We expect that we are able 
to use $N_\mu = 10^{21}/$year, and the total running time to be of the order of an year.
%
%
For the measurement of $P(\nu_e \to \nu_\mu)$, we need to distinguish
$\nu_\mu$ from $\bar \nu_\mu$ at Hyper-Kamiokande.
The water Cherenkov detector such as Hyper-Kamiokande can identify the charge of the muons generated by the charged current process~\cite{Beacom:2003nk,Huber:2008yx} by tagging the neutron in the final state, but the efficiencies of 
such signals are currently under
studies~\cite{Beacom:2003nk,Super-Kamiokande:2023xup,Akutsu:2019}.
We define the oscillation probability as follows.
\begin{align}
    &\scalebox{0.9}{$P (\nu_e\to\nu_\mu)=$} 
    \notag \\
    &\scalebox{0.9}{${1 \over \kappa} \frac{\qty(\kappa N_{\rm far}^{\nu_e\to\nu_\mu}+(1-\kappa)N_{\rm far}^{\bar{\nu}_\mu\to\bar{\nu}_\mu})-(1-\kappa)N_{\rm far}^{\bar{\nu}_\mu\to\bar{\nu}_\mu} \big|_{\rm T2HK}}{\tilde{N}_{\rm near}^{\nu_e\to\nu_e}}$}
    \label{eq10}
\end{align}
and
\begin{align}
    \kappa &\equiv \frac{1+C_{\rm id}}{2}.
    \label{eq11}
\end{align}
Here, $\tilde{N}_{\rm near}$ and $N_{\rm far}$ denote the number of events at the source and at the detector, respectively, while $\tilde{N}_{\rm near}$ is
corrected by the factors of the detector volume and distance. The factor $\kappa$ represents
the probability to correctly identify $\nu_\mu$ for an actual
$\nu_\mu$ event.
We assume that $N_{\rm far}^{\bar \nu_\mu \to \bar
\nu_\mu}\big|_{\rm T2HK}$ can be estimated by using T2HK data.
Combining $P(\nu_e \to \nu_\mu)$ defined in Eq.~\eqref{eq10} with  $P(\nu_\mu \to \nu_e)$ measured at the T2HK
experiment, one can now obtain the T-violation for
each energy bin labeled by $j$ such as follows.
\begin{align}
    P_j^{\rm TV} = P_j(\nu_e - \nu_\mu) 
    - P_j (\nu_\mu - \nu_e) \big|_{\rm T2HK}
    \label{eq13}
\end{align}
In defining $\chi^2$, we take the CP phase $\delta$ and the
matter density $\rho$ as the input parameters, and discuss how well we
can measure $\delta$. We simply assume that we do not know the actual value. 
For the rest of oscillation parameters, we take the following
reference values. In this study, we consider only the case of NO (Normal Ordering) for illustration.
\begin{table}[H]
    \centering
    \begin{tabular}{c|c}
         $\mathit{\Delta} m_{21}^2/10^{-5}\ \mathrm{eV}$ 
         &$\mathit{\Delta} m_{31}^2/10^{-3}\ \mathrm{eV}$ 
            \\
         \hline 
         $7.43$ & $2.432$ \\
    \end{tabular}
\end{table}
\begin{table}[H]
    \centering
    \begin{tabular}{c|c|c}
         $\theta_{12}$
         &$\theta_{13}$
         &$\theta_{23}$ \\
         \hline 
         $\ang{33.9}$ & $\ang{8.49}$ & $\ang{48.1}$ \\
    \end{tabular}
    \caption{The reference values of oscillation parameters for the normal mass ordering.}
    \label{tab1}
\end{table}
When we take the true values for the input parameters as $\delta_0$
and $\rho_0$, $\chi^2$ for the T-violation measurement as a function
of postulated values of $\delta = \delta^{\rm test}$ and $\rho =
\rho^{\rm test}$ is defined as
\begin{align}
    &\scalebox{0.9}{$\chi^2_{\rm TV} (\delta^{\rm test}, \rho^{\rm test})$}
    \notag\\
    &\scalebox{0.9}{$\equiv \sum_j \frac{\qty[ P^{\rm TV}_j\qty(\delta_0,\ \rho_0) - P^{\rm TV}_j\qty(\delta^{\rm test},\ \rho^{\rm test}) ]^2}{\qty(\mathit{\Delta}P^{T\rm V}_j)^2}$} \label{eq12} 
\end{align}

\begin{figure}[H]
    \centering
    \includegraphics[width=7.5cm]{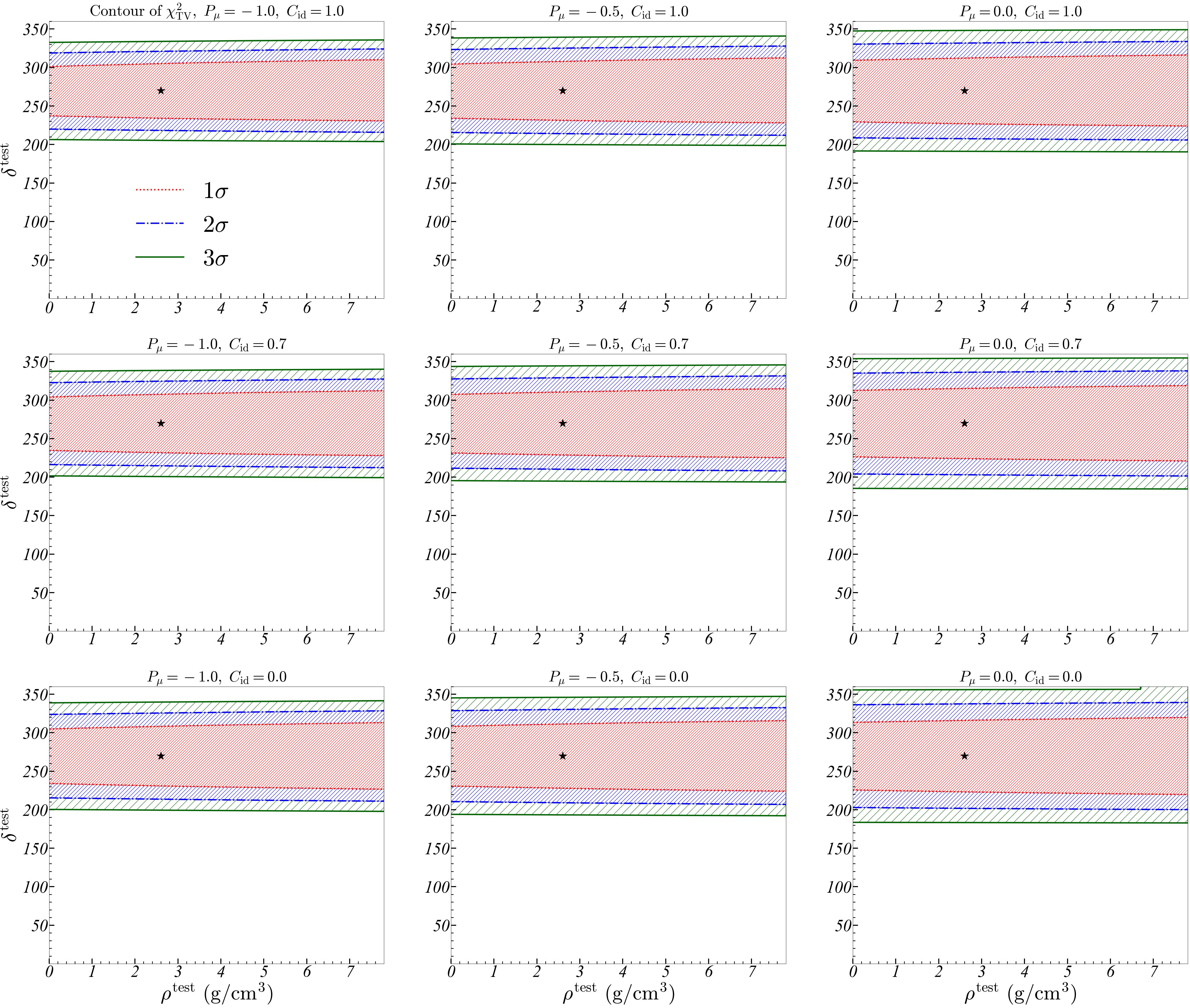}
    \caption{The contour of $\chi^2_{\rm TV}$ is defined in Eq.~(\ref{eq12}). The shaded regions represent $1\sigma$, $2\sigma$, and $3\sigma$ confidence interval, respectively, from inside to outside. From top to bottom, the figures show the cases with $C_{\rm id}=1.0,\ 0.7,\ 0.0$. From left to right, the figures show the cases with the anti-muon polarization $P_\mu = -1.0,\ -0.5,\ 0.0$. The stars in these figures represent the true input value ($\delta_0=-\pi/2,\ \rho_0=2.6\ \mathrm{g/cm^3}$).}
    \label{fig11}
\end{figure}

\begin{figure}[H]
    \centering
    \includegraphics[width=6.5cm]{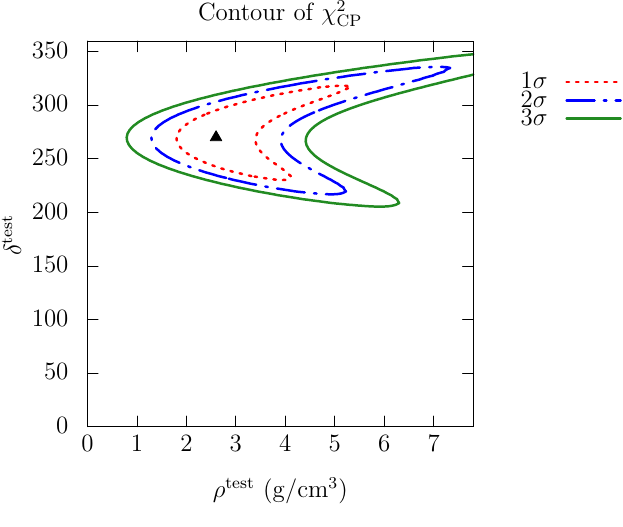}
    \caption{The contour of $\chi^2_{\rm CP}$. The red dot, blue
    dashdot, and green solid lines represent $1\sigma$, $2\sigma$, and
    $3\sigma$ confidence interval, respectively. The triangle symbol represent the true input value
    ($\delta_0=-\pi/2,\ \rho_0=2.6\ \mathrm{g/cm^3}$).}
    \label{fig10}
\end{figure}

In Fig.~\ref{fig11}, we show the contour plots of $\chi^2_{\rm TV}
(\delta^{\rm test}, \rho^{\rm test})$.
The results are almost unchanged for different choices of $C_{\rm id}$.
Independent of $C_{\rm id}$, CP (or T) conserving point, $\delta = 0$ and
$\pi$ can be excluded at the level of 3$\sigma$ even for the
unpolarized muon beam.
For the best case, $C_{\rm id} = 1.0$ and $P_\mu = -1.0$, the
CP phase $\delta$ is determined with an accuracy of about $\pm
\ang{30}$.
It is clear from the figure that $\chi^2_{\rm TV}$ depend on
$\rho^{\rm test}$ only very slightly. 
In contrast to the T-violation measurement, CP violation suffers from the uncertainties in the matter profile of the earth.
We consider CP violation,
\begin{align}
    \scalebox{0.9}{$P^{\rm CP}_j = P_j (\nu_\mu \to \nu_e) \big|_{\rm T2HK} 
    - P_j (\bar \nu_\mu \to \bar \nu_e) \big|_{\rm T2HK}
    $}
\end{align}
and we define $\chi^2_{\rm CP}
(\delta^{\rm test}, \rho^{\rm test})$ in a similar way as T violation. 
We show the contour of the
confidence interval in Fig.~\ref{fig10}.
We see a non-trivial dependence on $\rho^{\rm test}$, and it is clear that a good knowledge of the matter density
profile will be necessary. The measurement of
T violation will be an important additional information for the
measurement of $\delta$. 
%
Although we do not try in this paper, 
one would be able to perform a similar analysis for CPT violation,
\begin{align}
    \scalebox{0.9}{$P_j^{\rm CPT} =$} &\scalebox{0.9}{$P_j (\nu_e \to \nu_\mu)$} 
    \scalebox{0.9}{$-P_j (\bar \nu_\mu \to \bar \nu_e) \big|_{\rm T2HK} $}
    \notag \\
    \scalebox{0.9}{$=$}& \scalebox{0.9}{$P_j^{\rm TV} + P_j^{\rm CP}$}
    \label{eq:cpt}
\end{align}
The points of measuring/observing T violation in future 
experiments are, however,   
an independent crosscheck of the ``previous'' results from T2HK with different systematics and more importantly a test of the CPT theorem by separately measuring the two terms in Eq.~\eqref{eq:cpt}, rather than improvement of the statistical precision to measure $\delta$.

Under reasonable assumptions on the muon intensity, the results show that the observation of T violation is possible. 
For CP violation, $\chi^2_{\rm CP}$ depends significantly on both $\delta$ and $\rho$, whereas for T violation, $\chi^2_{\rm TV}$ is almost independent on $\rho$. T violation would not suffer from the uncertainties in the matter density profile of the earth. In this study, we considered only the
statistical error. A more complete analysis will be necessary to
establish the feasibility.
This proceeding is based on JHEP \textbf{12} (2024), 014 [arXiv:hep-ph/2407.05807].

\section*{Acknowledgements}
We would like to thank Osamu Yasuda for useful discussions and 
Ken Sakashita and Takasumi Maruyama for useful
information on neutrino detectors. This work was in part supported by
JSPS KAKENHI Grant Numbers JP22K21350 (R.K.), JP21H01086 (R.K.),
JP19H00689 (R.K.), and JST SPRING Japan Grant Number JPMJSP2178.
\bibliographystyle{./utphys.bst}
\bibliography{./T-violation.bib}
\end{multicols}
\end{document}